\documentclass[twocolumn,prl,aps]{revtex4}
\begin{document}
\title{Comment on ``Metallization of Fluid Nitrogen and the Mott 
Transition in Highly Compressed Low-Z Fluids''}
\author{Marina Bastea}
\affiliation{Lawrence Livermore National Laboratory, P. O. Box 808, 
Livermore, CA 94550}
\maketitle
The physical behavior and microscopic nature of highly compressed fluids 
continue to be unsolved and actively debated topics. The experimental data 
presented by Chau et al. on the electrical conductivity of nitrogen at 
high pressures are a welcome addition to the field. Unfortunately, in 
their desire to achieve a unifying picture of the behavior of low-Z fluids 
the authors oversimplify the interpretation of their data and {\it de facto} 
overlook the unique features of the systems they discuss. 

The main hypothesis of the ``low Z-systematics'' introduced by Chau et al. 
is that nitrogen, hydrogen and oxygen are completely dissociated in the 
shock reverberation experiments that lead to their metallization. This is 
hardly a reasonable assumption for oxygen. The results and analysis of 
electrical conductivity experiments presented in Ref. \cite{bastea-O2} 
indicate that metallization of fluid oxygen occurs in the molecular phase, 
and not in an atomic regime as stated in \cite{chau-N2}. This conclusion 
is supported by recent {\it ab initio} simulations of fluid oxygen at the 
conditions of the shock reverberation experiments \cite{galli-O2}. The 
simulations also reveal unique features of fluid oxygen under pressure 
e.g., the role of the triplet spin state in the metallization transition 
and an unusual behavior of the molecular bond \cite{galli-O2}.

Complete dissociation of nitrogen at $80 GPa$, the lowest pressure in the 
experiments of \cite{chau-N2}, is not in fact supported by previous 
work. The authors arguments are loosely based on two theoretical 
calculations but overlook their explicit caveats and disagreements 
with experiments. The QMD simulations invoked in \cite{chau-N2} report 
``noticeable disagreement from the second shock'' experiments in the 
dissociation region \cite{LANL-N2reshock}. It is therefore clear that 
any extrapolation of this work to the multiple-shock states relevant 
for the conductivity experiments has to be done with caution 
\cite{LANL-N2reshock}. As additional evidence for an atomic state 
assumption the authors use a dissociation energy calculated by Ross 
\cite{ross-N2} along with an empirical criterion introduced in 
\cite{nellis-h2}. However, by consistently following the treatment 
of \cite{ross-N2}, the first shock ($14 GPa$) and reshock ($41 GPa$) 
dissociation fractions  corresponding to the $80 GPa$ final pressure 
can be estimated to about $10^{-10}$ and $3 \times 10^{-3}$ respectively. 
Since temperature reaches over $90\%$ of its final value upon reshock 
it is reasonable to assume that the fluid retains a significant molecular 
component. After all even on the Hugoniot at $80 GPa$ and temperatures 
over $1.2eV$ in both theoretical studies the fluid appears to be 
$30-40\%$ molecular \cite{LANL-N2reshock,ross-N2}.

A discussion of the hydrogen metallization experiments and Mott scaling 
analysis can be found in \cite{nellis-96}, where evidence is presented 
in support of a molecular fluid, in contradiction with the approach taken 
in \cite{chau-N2}. 

We would also like to point out several other inconsistencies of
the ``low-Z fluids systematics'' presented in \cite{chau-N2}. For
example, we calculate a Mott scaling parameter of $.27$ for nitrogen at the
metallization conditions defined in the paper, in disagreement with the .35
value quoted by the authors. This comes simply from an average distance between
atoms of $1.81 \AA$ at a density of $3.9g/cm^3$, and a Bohr radius (location
of the maximum in the valence charge distribution $r^2\psi\psi^{\star}$
averaged over the solid angle) of $.91 bohr$, Ref. [26] of \cite{chau-N2}. 
The authors use of an empirical atomic radius meant to fit 
interatomic distances in a variety of crystalline compounds, see Ref. [25] of 
\cite{chau-N2}, is puzzling and unnecessary. For example the quoted reference 
gives a $.5 bohr$ radius for hydrogen, in large disagreement with the accepted 
$1 bohr$ value. It should also be noted that although using zero pressure 
charge distributions to gain insight into the behavior at high pressure is 
perhaps reasonable, shifting the curves in the radial direction as done in 
Fig. 3 of \cite{chau-N2} in order to compare their spatial extent is a rather 
meaningles exercise. Scaling all distributions to unity would allow the intended 
comparison and make it more apparent that the radial extent of hydrogen 
exceeds that of atomic nitrogen in contradiction with the statements by Chau 
et al.

Nitrogen is a complex system with one of the largest dissociation energies 
known in its diatomic state, a multitude of chemical bonding configurations 
and a high pressure polymeric phase \cite{hemN2-mcmah}. An accurate 
interpretation of electrical conductivity data for nitrogen around its high 
pressure metallization transition will probably need to account for this 
complexity. 

This work was performed under the auspices of the U. S. Department of Energy by 
University of California Lawrence Livermore National Laboratory under Contract 
No. W-7405-Eng-48.


\begin{thebibliography}{99}
\bibitem{chau-N2} R. Chau et al, Phys. Rev. Lett. {\bf 90}, 245501 (2003).
\bibitem{bastea-O2} M. Bastea et al, Phys. Rev. Lett. {\bf 86}, 3108 (2001).
\bibitem{galli-O2} B. Millitzer et al, to appear in Phys. Rev. Lett. (2003).
\bibitem{LANL-N2reshock} S. Mazevet et al, Phys. Rev. B {\bf 65}, 014204 (2001).
\bibitem{ross-N2} M. Ross, J. Chem. Phys. {\bf 86}, 7110 (1987).
\bibitem{nellis-h2} W.J.Nellis Phys. Rev. Lett. {\bf 89}, 165502 (2002).
\bibitem{nellis-96} S.T. Weir et al Phys. Rev. Lett. {\bf 76}, 1860 
(1996); W.J. Nellis et al Phys. Rev. B {\bf 59}, 3434 (1999).
\bibitem{hemN2-mcmah} M.I. Eremets et al, Nature {\bf 411}, 170 (2001);
A.K.McMahan et al, Phys. Rev. Lett., {\bf 54}, 1929 (1985).
\end{thebibliography}
\end{document}